\documentclass[runningheads]{llncs}
\usepackage{graphicx}
\usepackage{amsfonts}
\usepackage{amsmath}
\usepackage{algorithm}
\usepackage{algpseudocode}
\usepackage{multicol}
\usepackage{cite}
\usepackage{fancyvrb}
\usepackage{textcomp}

\begin{document}

\title{Efficient Verification of Optimized Code}
\subtitle{Correct High-speed X25519}
\titlerunning{Correct High-speed X25519}
\author{Marc Schoolderman\inst{1,2} \and
Jonathan Moerman\inst{1} \and
\\
Sjaak Smetsers\inst{1} \and
Marko van Eekelen\inst{1,2}}
\authorrunning{M.R. Schoolderman et al.}
\institute{
Radboud University, Nijmegen, The Netherlands\\
\email{\{m.schoolderman, jmoerman, s.smetsers, marko\}@science.ru.nl}
\and
Open University of the Netherlands, Heerlen, The Netherlands\\
\email{\{marc.schoolderman, marko.vaneekelen\}@ou.nl}
}

\maketitle

\begin{abstract}
Code that is highly optimized poses a problem for program-level verification:
programmers can employ various clever tricks that are non-trivial to reason about.
For cryptography on low-power devices, it is nonetheless crucial that implementations be functionally correct, secure, and efficient.
These are usually crafted in hand-optimized machine code that eschew conventional control flow as much as possible.

We have formally verified such code: a library which implements elliptic curve cryptography on 8-bit AVR microcontrollers.
The chosen implementation is the most efficient currently known for this microarchitecture.
It consists of over 3000 lines of assembly instructions.

Building on earlier work, we use the Why3 platform to model the code and prove verification conditions, using automated provers.

We expect the approach to be re-usable and adaptable, and it allows for validation. Furthermore,
an error in the original implementation was found and corrected, at the same time reducing its memory footprint.

This shows that practical verification of cutting-edge code is not only possible, but
can in fact add to its efficiency---and is clearly necessary.

\end{abstract}

\newcommand{\ratio}[2]{#1\!:\!#2}

\section{Introduction}
Although formal verification is considered to give the highest level of assurance in security-critical software\cite{ealcc},
it is seldom applied.
Even if a verification technique is expressive enough to reason about a given problem domain,
for its use to make economic sense,
it must be usable by programmers proficient in that domain,
and not require an excessive amount of time.
These criteria are hard to meet.

Cryptographic implementations are always security-critical: subtle bugs can have disastrous consequences\cite{bugattacks}, and the security of a system is only as strong as its
weakest link.
As Chen et al.\cite{cv25519} note, the desire to avoid risk in cryptographic implementations can hamper adoption of new 
and more efficient crypto libraries, simply because the correctness of these implementations cannot be properly demonstrated. 
As they also note, a full audit in addition to testing can be extremely expensive, and impractical for
high-performance implementations due to extensive use of clever optimizations. In this context the case of applying formal
verification looks very reasonable, and indeed this is actively pursued \cite{cv25519, everest}.

However, this poses many important challenges. First, at what level should verification occur? Compilers
have been known to be a source of concern, as they can cause subtle problems\cite{vaudenay}. Second, understanding
the formal verification process used can be a daunting task: powerful tools such as the Verification Software
Toolchain \cite{vst} have a substantial learning curve. If instead an ad-hoc method is used, the correctness of
the method itself needs to be clearly established for it to be trustworthy.
Lastly, cryptography by its nature involves the exploitation of carefully engineered mathematics, which a formal method
must be able to state and work with, which adds to the effort required in showing correctness of implementations.

To rely on the verification of any code---cryptographic or otherwise---its specification must be
validated as well. This demands a formal specification that is succinct, and comprehensible by a domain expert.
Furthermore, we do not want to decide between efficiency and correctness: both are important, and in fact verification
ideally assists in making implementations more efficient. Finally, for a verification technique to be practical, it
should be re-usable for other verification tasks in the future, and not simply a one-shot operation.

In this paper, we present such a technique, by applying the existing Why3 verification
platform\cite{why3} to prove the functional correctness of a highly optimized library used for
X25519 elliptic curve cryptography on 8-bit microcontrollers\cite{hutter}. We arrive at a succinct specification, and we expect our technique
to be capable of verifying similar code for more powerful processors with less effort.

\subsection{Contributions}
We provide a corrected version of an X25519 implementation optimized for the 8-bit AVR architecture. Our modifications, described in Section~\ref{sec:bugs}, improve
upon the fastest implementation currently known for this challenging architecture\cite{hutter}.

We demonstrate functional correctness and memory safety of this implementation by providing a machine-checked proof 
using the Why3
verification platform\cite{why3}. Concretely, we prove that the code calculates a scalar multiplication on Curve25519 by applying a double-and-add scheme---the \emph{Montgomery ladder}---using Montgomery's $x$-coordinate-only formulas\cite{monty87}.

We also provide a formal Why3 model of a subset of the AVR instruction set, that has been carefully constructed for easy
validation with respect to the official specification\cite{avrref}. This model can be re-used for other purposes, or
modified to fit a different verification purpose without loss of its validity.

We describe our approach in using Why3 for this verification task; this is an extension of earlier work\cite{vstte17}, and has
been demonstrated to have a low barrier to understanding\cite{cserc19}. This approach should work similarly well for
other architectures such as ARM or RISC-V. The overall methodology is not specific to the domain of cryptographic implementations.

\subsection{Availability of Results}
The code belonging to this paper is available online
in an open repository.\footnote{\url{https://doi.org/10.5281/zenodo.4640377}} To
check the proofs, Why3 version 0.88.3 is required.\footnote{
Later versions do not yet support our approach---see Section~\ref{sec:conclude}
}
For discharging the verification conditions the provers CVC3 (2.4.1), CVC4 (1.4 and 1.6), Z3 (4.6.0), and E-prover (2.0) were used.

\section{Elliptic Curve Cryptography on Small Devices}
\label{sec:overview}
X25519 is a public key cryptography scheme built around
a Diffie-Hellman key exchange\cite{bernstein-dh,rfc7748}. `Original' Diffie-Hellman obtains its 
security through the observation that, given a primitive root $g$ for a prime $p$,
it is (in general) hard to compute $g^{xy}$ from $g^x$ and $g^y$ (mod $p$) without knowing the integers $x$ or $y$\cite{dh-dlp}.
For proper security, a sufficiently large prime modulus $p$ is
needed---2048 bits is a recommended minimum\cite{rfc8270}. Performing the required exponentiation
and modular reduction steps on such large integers is hard to do efficiently on restricted devices\cite{gura2004}. Also,
the viability of side-channel attacks prescribes various precautions on all code that computes
using secret data, to ensure that an implementation does not inadvertently leak information\cite{maythefourth}.

\subsection{Curve25519}
Using elliptic curves eases some of these issues\cite{bernstein-dh}. Given a field $\mathbb{F}$, and coefficients $A,B\in\mathbb{F}$,
a \emph{Montgomery curve over $\mathbb{F}$} is defined as all the points $x,y \in \mathbb{F}$ that satisfy the formula:
$$
By^2 = x^3 + Ax^2 + x
$$
To this set of points is added a `point at infinity' denoted $\mathcal{O}$ to form an additive group.
When $P, Q$ are \emph{distinct} points on the curve, $P+Q$ is defined as the third point on the curve that intersects the
straight line passing through $P$ and $Q$, reflected around the $x$-axis.
For $P+P$, the tangent of the curve at point $P$ is used to find this point. 
The point at infinity $\mathcal{O}$ acts as the neutral element.

The separate cases of point \emph{adding} and \emph{doubling}, can be used to compute a \emph{scalar multiple} $n\cdot P$, or $P$ added to itself $n$ times, using a double-and-add scheme. Again a \emph{Diffie-Hellman assumption}\cite{kleppmann} applies: if $\mathbb{F}$ is a finite field of prime order, it is assumed to be hard to compute $nm\cdot P$ from $n\cdot P$ and $m\cdot P$.

X25519 performs a scalar multiplication on Curve25519: a Montgomery curve over the finite 
field $\mathbb{F}_p$ where $p=2^{255}-19$, and coefficients $A=486662, B=1$. The choice of $p$ facilitates efficient modular
reductions. Furthermore, Montgomery\cite{monty87} gives efficient formulas for both \emph{doubling} and \emph{differential addition} of points, which only requires the $x$-coordinates of points. These formulas derive their efficiency by representing an $x$-coordinate by the ratio $\ratio{X}{Z}$, with $x \equiv X \cdot Z^{-1}~\text{mod}~p$.
\label{sec:ecc}

The scalar multiple $n\cdot P$, finally, is computed using the \emph{Montgomery ladder}. This can be mathematically described by the following formula:
\label{sec:mladder}
\begin{align*}
	\textsc{ladder}~n~P =\begin{cases}
		(\mathcal{O},~P) & \text{if $n = 0 $}\\
		(2R_0,~R_1+R_0) & \text{if $n > 0$ and even}\\
		(R_1+R_0,~2R_1) & \text{if $n > 0$ and odd}
	\end{cases}
\\
	 \textrm{where in the last two cases~} (R_0,~R_1) &= \textsc{ladder}~\lfloor n/2 \rfloor~P
	 \label{form:ladder}
\end{align*}
It can be shown that for every $n\ge 0$, $\textsc{ladder}~n~P = (n\cdot P,~(n+1)P)$, but instead of computing $n\cdot P$
using a naive double-and-add scheme, this definition performs the same arithmetic operations in both recursive cases --- the only difference between the recursive cases is a swap of the arguments. This enables a constant-time implementation\cite{bernsteinlange}.

\subsection{X25519 on AVR}
The AVR microarchitecture is an 8-bit RISC architecture\cite{avrref}, and so we can only represent
an element $x\in\mathbb{F}_p$ by splitting it into 32 bytes.
Since the AVR only has 32 registers (of which some are needed as index registers),
no single element $x\in\mathbb{F}_p$ can be loaded from memory entirely.
Therefore, judicious register allocation is of prime concern for an efficient
implementation.
Therefore, all of the primitives operations in $\mathbb{F}_p$ are rendered in assembly code in \cite{hutter}. These comprise the following:
\begin{itemize}
	\item A $256\to256$-bit routine subtracting $2^{255}-19$ from its input (with borrow).
	\item A $256\times256\to512$-bit multiplication routine, constructed by recursive application of Karatsuba's algorithm out of smaller $32\times32\to64$-bit multiplication routines.
	\item A $256\to512$-bit dedicated squaring routine of similar construction
	\item A $512\to256$-bit modular reduction function, which given a $m\in\mathbb{F}_p$ computes $\hat m$ so that $\hat m \equiv m~(\text{mod}~p)$ and $\hat m < 2^{256}$, used to reduce the results of the previous two functions.
	\item $256\times 256\to256$-bit modular addition/subtraction routines which perform a multi-precision addition/subtraction with a built-in modular reduction.
	\item A specialized $256\to256$-bit routine for efficient modular multiplication with the constant 121666.
\end{itemize}
Other operations are rendered in C code: these are either very simple, or consist mostly of function calls to these primitive operations. Examples of such functions would be a $256\times 256\to 256$-bit modular multiplication, a function that canonicalizes an element $x\in\mathbb{F}_p$ by repeated subtraction of $p$, and a function that computes $x^{-1}~(\text{mod}~p)$ using Fermat's little theorem. 

The Montgomery ladder is implemented in C iteratively as illustrated by Algorithm~\ref{alg:mladder}. Essentially this computes
the scalar multiple using the same double-and-add scheme as the \textsc{Ladder} function defined above, starting at the least significant bit of its input, and swapping the roles of
$(\ratio{X_1}{Z_1})$ and $(\ratio{X_2}{Z_2})$ as needed.
We will show in Section~\ref{sec:laddernote} that the informal specification given here is \emph{not} entirely correct.
The \textsc{Ladderstep} procedure shown in Algorithm~\ref{alg:ladderstep} is an optimized implementation of Montgomery's formulas\cite{monty87} for doubling and adding points. Note that the literature usually only presents the Montgomery ladder in this iterated version, often---confusingly---with minor variations to the \textsc{Ladderstep} procedure\cite{rfc7748, bernstein-dh}. We find this optimized form of the Montgomery ladder hard to understand, making its full formal verification desirable.

\begin{algorithm}[h]
	\caption{Montgomery ladder for scalar multiplication}
	\begin{algorithmic}[0]
		\Require A 255-bit scalar $n$, and a $x$-coordinate $x_P$ of a point $P$
		\Ensure Result $(\ratio{X}{Z})$ satisfies $x_{n\cdot P} \equiv X\cdot Z^{-1}$
		\State $(\ratio{X_1}{Z_1}) \gets (\ratio{1}{0})$; $(\ratio{X_2}{Z_2}) \gets (\ratio{x_P}{1})$;
		$prev \gets 0$; $j \gets 6$
		\For{$i \gets 31 ~\textbf{downto}~ 0$}
		\While{$j \ge 0$}
		\State $bit \gets \textbf{bit~} 8i+j \textbf{~of~}n$
		\State $swap \gets bit \oplus prev$; $prev \gets bit$
		\State \textbf{if~}swap\textbf{~then}
 		       $(\ratio{X_1}{Z_1},\ratio{X_2}{Z_2}) \gets (\ratio{X_2}{Z_2},\ratio{X_1}{Z_1})$
		\Comment{by conditional moves}
		\State $\textsc{Ladderstep}(x_P,X_1:Z_1,\ratio{X_2}{Z_2})$
		\State $j \gets j-1$
		\EndWhile
		\State $j \gets 7$
		\EndFor
		\State \Return $(\ratio{X_1}{Z_1})$
	\end{algorithmic}
	\begin{multicols}{2}
	\begin{algorithmic}[0]
		\Procedure{Ladderstep}{}
		\State $T_1 \gets X_2+Z_2$
		\State $X_2 \gets X_2-Z_2$
		\State $Z_2 \gets X_1+Z_1$
		\State $X_1 \gets X_1-Z_1$
		\State $T_1 \gets T_1 \cdot X_1$
		\State $X_2 \gets X_2 \cdot Z_2$
		\State $Z_2 \gets (Z_2)^2$
		\State $X_1 \gets (X_1)^2$
		\State $T_2 \gets Z_2 - X_1$
		\State $Z_1 \gets T_2\cdot 121666$
		\State $Z_1 \gets Z_1 + X_1$
		\State $Z_1 \gets T_2 \cdot Z_1$
		\State $X_1 \gets Z_2 \cdot X_1$
		\State $Z_2 \gets T_1 - X_2$
		\State $Z_2 \gets (Z_2)^2$
		\State $Z_2 \gets Z_2\cdot x_P$
		\State $X_2 \gets T_1 + X_2$
		\State $X_2 \gets (X_2)^2$
		\EndProcedure
	\end{algorithmic}
	\end{multicols}
	\label{alg:mladder}
	\label{alg:ladderstep}
\end{algorithm}

\section{Why3 Verification Platform}
\label{sec:why3}
Why3\cite{why3} is a verification platform for deductive program verification. It comprises the typed programming
language WhyML (which can be annotated with functional contracts and assertions),
as well as libraries for reasoning about specific types of objects (such as arrays, bit-vectors, bounded and unbounded integers), which the
user can also extend.
A weakest-precondition calculus generates the correctness condition for an annotated program,
which Why3 then transforms into the input language for various automated or interactive provers.
Besides assertions and contracts, WhyML also provides other means of instrumenting programs to aid verification. We highlight two:

\begin{description}
	\item[Ghost code] is guaranteed by the type system to not have any effect on the actual execution on the code, but can be used to compute witnesses for use in verification goals.
	\item[Abstract blocks] can be used to summarize multiple operations with a single functional contract.
\end{description}

An advantage of Why3's reliance on automatic provers is that verification does not need to be the last step in a waterfall-like process.
When a program (or specification) is changed, most of the verification conditions that held previously
can usually be solved again at the press of a button, even when the change affects them.
Similarly, if a prover can solve one instance of a problem, it can usually---given enough time---handle similar or larger instances, allowing for proofs to be transplanted. For instance, we recycled parts of the proofs of \cite{vstte17}.
Since Why3 uses multiple provers in concert, we are not restricted by the limitations of one particular (version of) a prover.
In this sense, proofs seem robust.

On the other hand, too much irrelevant information can hinder automatic provers.
Sometimes an assertion frustrates a proof that is completely unrelated to it.
In this sense, proofs can also be brittle. Thus, for large verification tasks keeping the proof context small is vitally important. We used Why3's module system, \emph{ghost code} and \emph{abstract blocks} to keep the proof context manageable.

\section{Correctness of Low-level Code}
\label{sec:asm}
In the implementation we are considering, all primitives for implementing the field arithmetic needed for computing in $\mathbb{F}_p$ are
implemented in assembly code. With the exception of the multiplication
routine, this code is free of conditional branches. In the multiplication routine, branches are used,
but in every case, both branches take the same amount of clock cycles and perform the same sequence of memory accesses. This should prevent a side-channel attack such as described by Genkin et al.\cite{maythefourth}, which exploits observed timing differences.
Our formal verification effort therefore only focuses on the functional correctness and memory-safety of these routines.

Since $256$-bit operations are not natively supported on any CPU, an X25519 implementation usually chooses a
representation where
an element $x\in\mathbb{F}_p$ is represented in $n$ \emph{limbs} in radix $2^w$; that is, 
$x = \sum^n_{i=0} 2^{iw} x_{[i]}$ for the limbs $x_{[0]}, x_{[1]}, \ldots x_{[n-1]}$. If these limbs can contain
more than $w$ bits of information, this representation is called \emph{unpacked}, and any carry that occurs during
computation does not need to be propagated to the next limb immediately. An \emph{unpacked} representation
with few limbs is more efficient, and is thought to be more convenient for verification\cite{cv25519, zinzin}.
On the implementation for the AVR a \emph{packed} representation of 32 limbs in radix $2^8$ is used,
and every part of the code is forced to handle carry-propagation.

Globally, our approach follows that of \cite{vstte17}; we specify the representation of a $256$-bit multi-precision integer
in terms of an 8-bit memory model, model every AVR mnemonic that is needed as a WhyML function, and mechanically translate the
assembly code to this model for verification with Why3.

\subsection{A Re-usable Validated AVR Machine Model}
\label{sec:avrmodel}
For modeling the processor state, we use the concept of an \emph{8-bit address space}, which is a Why3 \texttt{map} of addresses to integers, suitably restricted:
\begin{verbatim}
type address_space = { mutable data: map int int }
  invariant { forall i. 0 <= self.data[i] < 256 }
\end{verbatim}
The AVR register file, data segment, and stack are all modeled as separate address spaces.
This of course means that our model is an underspecification, but most
assembly code conforms to this simplified model.
Memory size restrictions are not part of the definition of an \emph{address space}, as it is more convenient to express them as pre-conditions for the AVR instructions that manipulate memory.
To model the carry and `bit transfer' CPU flags, we use the equivalent of a
\texttt{ref bool};  the value of all other flags are unspecified. We also use \emph{ghost registers} \cite{vstte17} to
track register updates inside abstract blocks using Why3's type system.

Since we needed to model many AVR instructions, we first implemented (in WhyML) a \emph{primitive instruction set} of
common operations on these \emph{address spaces}, such as reading and writing 8-bit and 16-bit values represented either by
their integer value, or as bit-vectors. These operations are verified for consistency with the \emph{8-bit address space}.
This instruction set is then used to \emph{implement} all required AVR instructions
following the official specification\cite{avrref}. 

For example, for the \texttt{SUBI} instruction, the AVR specification tells us that a constant $K$ will be 
subtracted from its destination register, and the carry flag will be set to $\overline{r_7}\cdot K_7 + K_7 \cdot r'_7 + r'_7 \cdot\overline{r_7}$ (in boolean arithmetic), where $x_7$ denotes the most significant bit of an 8-bit value $x$, and $r, r'$ are the previous and updated values of the
destination register, respectively. In terms of our primitives, we can state this as:
\begin{Verbatim}
let subi (rd: register) (k: int)
  requires { 0 <= k <= 255 }
= let rdv  = read_byte reg rd in
  let res  = clip (rdv - k) in
  set_byte reg rd res;
  cf.value <- (not ar_nth rdv 7 && ar_nth k 7 ||
               ar_nth k 7 && ar_nth res 7 ||
               ar_nth res 7 && not ar_nth rdv 7)
\end{Verbatim}
While this follows the official specification closely, it is not very useful for verifying programs. 
Capturing the common notion that the carry flag gets set if and only if $r < K$ can be done by adding a Why3 contract for \texttt{subi}:
\begin{verbatim}
  ensures { reg = old reg[rd <- mod (old (reg[rd] - k)) 256] }
  ensures { ?cf = -div (old (reg[rd] - k)) 256 }
\end{verbatim}
That is, the register file gets updated with the destination register receiving $(r-K)~\textrm{mod}~256$, and the numeric value of the carry flag will be $-\lfloor \frac{r-K}{256} \rfloor$.

Why3 allows us to verify that this contract is satisfied by the AVR specification.\footnote{For \texttt{SUBI}, this also revealed a mistake in online documentation.}
Also, if a different contract were discovered to be more useful, it could easily be replaced while maintaining validity of the model.

\subsubsection{Extensions to the model}
Some of the code verified featured a limited form of branching.
We modeled this using a WhyML function that throws an exception if the branch is taken; this exception is then handled 
at the appropriate location.

In two locations, data on the stack was allocated for use with memory operations, which our simplified model did not support.
We resolved this by adding the requirement that the stack pointer does not alias with any of the ordinary data inputs, and checking
manually whether the code conforms to the conventions for accessing memory on the stack. As we will explain in 
Section~\ref{sec:interrupt}, this turned out not to be the case, necessitating modifications.

\subsection{Proving the correctness of AVR assembly code}
For all of the assembly routines, we of course want to show \emph{functional correctness}.
However, since these routines must interface with C code, we also have to verify that they are well-behaved. This means proving that they only modify the memory that they are allowed to (i.e. temporary data on the stack or that passed by
the caller as a pointer), that they leave the stack in a consistent state, and that they adhere to the C calling convention for
the AVR\cite{avrgcc}.

Note that there are two versions of the $256$-bit multiplication routines in\cite{hutter}: one which uses 
function calls to the
respective $128$-bit operation, and one which inlines everything for a very minor increase in speed. We consider the former
to be the more relevant one, and so have chosen that as our verification target. 

Quantitative verification results are shown in Table~\ref{tab:asmbench}.
The vast majority of the goals were discharged by CVC3 and CVC4.
The number of annotations required gives a
\emph{rough} measure of the manual effort. This is a subjective number since not
every annotation represents the same amount of effort.
As a point of reference, verifying \texttt{fe25519\_mul121666} was measured to take 16 hours of work.

\begin{table}[ht]
    \centering
\begin{tabular}{l|r|r|r|r}
	\emph{function} & \emph{instructions} & \emph{user annotations} & \emph{generated goals} & \emph{CPU time} \\
	\hline
	\texttt{bigint\_mul256:mul128} & 1078 & 122 & 300 & 1504.6s \\ 
	\texttt{bigint\_mul256} & 693 & 85 & 506 & 2000.1s \\ 
	\texttt{bigint\_square256:sqr128} & 672 & 26 & 135 & 363.8s \\
	\texttt{bigint\_square256} & 493 & 38 & 359 & 1796.6s \\
	\texttt{bigint\_subp} & 103 & 12 & 84 & 184.0s \\ 
	\texttt{fe25519\_red} & 305 & 41 & 182 & 155.3s \\
	\texttt{fe25519\_add} & 242 & 52 & 209 & 156.4s \\
	\texttt{fe25519\_sub} & 242 & 53 & 212 & 119.6s \\
	\texttt{fe25519\_mul121666} & 138 & 56 & 149 & 393.0s \\ 
\end{tabular}
	\caption{Results of verifying the X25519 field arithmetic in AVR assembly}
\label{tab:asmbench}
\end{table}

\subsubsection{Verification by partitioning into blocks}
The $256\times256$-bit multiplication is constructed by using calls to a $128\times 128$-bit multiplication routine using Karatsuba's method. The $128\times 128$-bit multiplication routine itself, is comprised of three in-line applications of a $64\times 64$-bit Karatsuba multiplication, the basic version of which was verified earlier in \cite{vstte17}. Some parts of this earlier proof could in fact simply be re-used. 

For the $128$-bit and $256$-bit larger versions, the proofs followed a similar approach, with one notable change. For the smaller Karatsuba routines, it sufficed to identify 7 `blocks' of code, and state their operations in \emph{contextual terms}---i.e., specifying which part of Karatsuba's algorithm each block performed.
For more than one level of Karatsuba, this becomes unwieldy. While we kept the identified blocks the same, we
found it much more useful---even for routines verified in \cite{vstte17}---to specify
their effects in purely \emph{local} terms---i.e, only specifying what its effect is in terms of its immediately
preceding state.
For some blocks, this simplifies the specification, and actually makes the work for automatic provers slightly easier.
In cases where this contextual information \emph{is} required, it can always be re-asserted later.
The only drawback we have found to this method was that on assembly code of this size, it is easy to lose sight of what one is trying to achieve without reliable contextual information.

The $256$-bit squaring routine is similarly constructed out of calls to a $128$-bit squaring routine;
both compute the square of $A = 2^w A_h + A_l$ as
$A^2 = (2^w+1) (2^w A_h^2 + A_l^2) - 2^w (A_l-A_h)^2$, which we are able to verify by partitioning these
routines into 5 blocks.

\subsubsection{Instrumenting programs with ghost code}
The routines that perform modular arithmetic are very different in style from the multiplication routines. In the latter,
we can apply a decomposition into a small number of large blocks, which allows SMT solvers to do most of the work. 
The reduction, addition and subtraction routines, by contrast, are highly repetitive---essentially the same read-modify-write sequence repeated several times.

In this case, it was more logical to use a bottom-up approach, summarizing the effects of these short sequences using
a WhyML function (essentially the same idea as using an assembly \emph{macro}), which is then iterated.
We discovered, however, that after a few macro applications, SMT solvers were unable to prove memory safety or
absence of aliasing. The culprit here seemed to be that the macros accessed memory via
\texttt{LD+}/\texttt{ST+} instructions (which perform a load/store, followed by a pointer increment).
Perhaps unsurprisingly, it becomes increasingly hard for SMT solvers to reason about out what address an
index register is referring to after many modifications have been applied to it.

In our routines (and we suspect, commonly in similar cases) such addresses are however perfectly obvious, and can be statically deduced. We therefore instrumented the code with \emph{ghost arguments}, which supply this missing information. 
As a simple example of this technique (which was also used in the $256\times256$-bit multiplication routine), we can make the
AVR \texttt{LD+} instruction (modeled as the WhyML function \texttt{AVRint.ld\_inc}) more amenable to verification
by instrumenting it with ghost arguments:

\begin{Verbatim}[commandchars=\\\[\]]
let ld_inc' (dst src: register) (ghost addr: int)
  ... \emph[(* the specification of AVRint.ld_inc *)]
  requires { uint 2 reg src = addr }
= AVRint.ld_inc dst src
\end{Verbatim}

On the surface, this just appears to add a needless pre-condition; however, once this correlation
between \texttt{addr} and \texttt{uint 2 reg src} is established, SMT solvers can use this information to easily
deduce what address the index register used is referring to.

\section{Correctness of the C Code}
\label{sec:c}
The X25519 implementation we verify also consists of around 300 lines of C code, which interfaces directly with the
assembly routines verified in Section~\ref{sec:asm}.
Many routines are short and simple, and verification for them is a straight-forward application of Why3.

To ensure that the C code and the assembly code are both verified with respect to the same logical foundations, we translate C
by hand into the WhyML primitives from Section~\ref{sec:avrmodel}, that underpin the AVR instruction set model.
However, since a C compiler handles allocation of global and local variables, using one \texttt{address\_space} to model memory
would be impractical and incorrect, as it would force the model to make assumptions about the memory layout. So instead, every
array object is modeled as residing in its own \texttt{address\_space}.
An added benefit of this is that Why3's type system will enforce that arguments do not alias.
The minor drawback is that some functions can be called to perform in-place updates, which does requires aliasing. These functions have to be modeled and verified for both cases separately.

For the assembly routines that interface with the C code, abstract specifications are added by duplicating
the contracts of the verified assembly routines, and removing the pre- and post-conditions related to the
C calling conventions.

The verification results are listed in Table~\ref{tab:cbench}. Among the field operations, it is notable that
\texttt{fe25519\_unpack} and \texttt{fe25519\_invert} generate more goals. The former is due to its (RFC-required)
bit-masking of its input, which we specify as a reduction $\text{mod}~2^{255}$.
We suspect our proof of this function can be further optimized, but decided against spending time on this.
Note that the field arithmetic code actually operates on a \emph{packed} representation, so 
\texttt{unpack} and \texttt{pack} functions are otherwise simply copy-operations.

The \texttt{fe25519\_invert} function computes $x^{2^{255}-21}~(\text{mod}~p)$ using sequences of modular square-and-multiply
steps.
This makes it very similar to repetitive
assembly code, and it is treated the same way: we instrument the code with \emph{ghost arguments} in a highly regular fashion
which specify the actual value of intermediate results---which interestingly was more or less a formalization of the
\emph{inline comments} provided by the original authors. Also, \emph{abstract blocks} helped keep the number of verification
conditions small. 

For the verification of the last three routines in Table~\ref{tab:cbench}, verification was `simply' an effort of finding the correct invariants and assertions that guided the automatic provers to the desired conclusion within an acceptable amount
of CPU time. To achieve the final conclusion presented in Section~\ref{sec:waarhetallemaalomtedoenwas}, it is required to know that $2^{255-19}$ is a prime number; we took the pragmatic route and stated this as an axiom in Why3.
To see that $x^{2^{255}-21}$ is the multiplicative inverse of $x$ also requires Fermat's little theorem, which we instead proved inside Why3 using \emph{ghost code} that traces a direct proof using modular arithmetic---showing that for any integer $a$ not divisible by a prime $p$, it is the case that
$a^{p-1}\prod^{p-1}_{i=1} i \equiv \prod^{p-1}_{i=1} a\cdot i \equiv \prod^{p-1}_{i=1} i$, and therefore $a^{p-1} \equiv 1$.

\begin{table}[ht]
    \centering
\begin{tabular}{l|r|r|r|r}
	\emph{function} & \emph{lines} & \emph{user annotations} & \emph{generated goals} & \emph{CPU time} \\
	\hline
	\texttt{fe25519\_setzero} & 3 & 2 & 7 & 0.4s \\
	\texttt{fe25519\_setone} & 4 & 2 & 7 & 0.4s \\
	\texttt{fe25519\_neg} & 3 & 0 & 3 & 0.2s \\
	\texttt{fe25519\_cmov} & 5 & 3 &10& 36.5s \\
	\texttt{fe25519\_freeze} & 4 & 2 &9& 4.7s \\
	\texttt{fe25519\_unpack} & 4 & 8 &30& 41.0s \\
	\texttt{fe25519\_pack} & 5 & 2 &11& 1.6s  \\
	\texttt{fe25519\_mul} & 3 & 0 &1& 0.2s   \\
	\texttt{fe25519\_square} & 3 &0&1& 0.1s \\
	\texttt{fe25519\_invert} & 51 &49 &306& 557.3s \\   
	\texttt{work\_cswap} & 8 &0&13& 3.8s \\
	\texttt{ladderstep} & 26 &22&80& 202.8s \\
	\texttt{mladder} & 26 &22&140& 345.1s \\        
	\texttt{crypto\_scalar\_mult\_curve25519} & 13 &27&57& 74.2s \\  
\end{tabular}
	\caption{Results of verifying the X25519 C routines}
\label{tab:cbench}
\end{table}

\subsection{Verifying the Montgomery Ladder}
Montgomery\cite{monty87} provides formulas for doubling and differential addition of
points on an elliptic curve, where only the $x$-coordinates of these points on the curve are used. As mentioned
in Section~\ref{sec:ecc}, these $x$-coordinates are represented as \emph{ratios} $(\ratio{X}{Z})$, where
$x \equiv X \cdot Z^{-1}~\text{mod}~p$.
The 
point at infinity $\mathcal{O}$, which is not on the curve, is represented by $(\ratio{X}{Z})$ with $X\ne 0, Z=0$.
The degenerate case $(\ratio{0}{0})$ does not represent anything.

For Curve25519, Montgomery's formulas are proven correct for all
cases by Bernstein\cite{bernstein-dh}, and look as follows:
\begin{align*}
	X_{2n}  &= (X_n^2 - Z_n^2)^2 &
	X_{m+n} &= 4 Z_{m-n} (X_m X_n - Z_m Z_n)^2 \\
	Z_{2n} &= 4 X_n Z_n (X_n^2 + 486662 X_n Z_n + Z_n^2) &
	Z_{m+n} &= 4 X_{m-n} (X_m Z_n - Z_m X_n)^2 \\
\end{align*}
If the $x$-coordinate of the point $n P$ is the ratio $(\ratio{X_n}{Z_n})$, then
$(\ratio{X_{2n}}{Z_{2n}})$ is the ratio for the point $(2n)P$.
Likewise, from $x_{nP}$ and $x_{mP}$, we can compute $x_{(m+n)P}$ provided we also know $x_{(m-n)P}$.

We have proven that the \texttt{ladderstep} procedure (see Algorithm~\ref{alg:ladderstep}),
given values $(x, \ratio{X_n}{Z_n}, \ratio{X_m}{Z_m})$, computes $(\ratio{X_{2n}}{Z_{2n}},
\ratio{X_{m+n}}{Z_{m+n}})$ as specified by these point doubling and addition
formulas, with $X_{m-n} = x$, and $Z_{m-n} = 1$.

To verify the function \texttt{mladder} (Algorithm~\ref{alg:mladder}),
we define a formal specification in Why3 of the Montgomery ladder as presented in Section~\ref{sec:mladder}, but using
the above formulas for doubling and addition. We verify that \texttt{mladder} adheres to this specification: if
for some 255-bit integer $s$ and $x$-coordinate $x_P$, 
$\textsc{ladder}~s~(\ratio{x_P}{1})$ returns $(\ratio{X}{Z})$ as the first component of its result,
\texttt{mladder} computes $(\ratio{\tilde X}{\tilde Z})$ such that $\tilde X\equiv X$ and $\tilde Z\equiv Z~(\text{mod}~p)$.

Importantly, for this result to hold, we found it necessary to require that $s$ is even, and has its most
significant bit set. The former is necessary, as an odd $s$ would leave the results of Algorithm~\ref{alg:mladder} in a state where a final swap is still needed.
Having bit $254$ in $s$ set is necessary, as it prevents Algorithm~\ref{alg:mladder} from performing the doubling formula on the `point at infinity', which would make it impossible to demonstrate the strict correspondence.

\label{sec:laddernote}
These requirements on $s$ are however taken care of by the existence of the `clamping' operation in X25519, which requires $s \in \{ 2^{254} + 8k : 0 \le k < 2^{251} \}$. Having $s$ a multiple of 8 is crucial for the mathematical security of X25519\cite{rfc7748}. Setting the high bit is done for entirely different reasons: to prevent programmers from applying a non-constant-time optimization that reveals information about the scalar $s$\cite{kleppmann}. Our formal proof was greatly helped by this choice, perhaps providing more justification for it.

\subsection{A Succinct Specification of X25519}
\label{sec:waarhetallemaalomtedoenwas}
The function \texttt{crypto\_scalar\_mult} is our ultimate verification goal.
We show the most important part of the specification proven in Why3 here:
\newcommand{\midtilde}{\raisebox{0.5ex}{\texttildelow}}
	\begin{Verbatim}[commandchars=\\\[\]]
val crypto_scalarmult_curve25519 (r s p: address_space)
  ensures { uint 32 r = mod (uint 32 r) p25519 }
  ensures { let xp   = mod (uint 32 p) (pow2 255) in
            let mult = scale (clamp (uint 32 s)) xp in
	    if mult \midtilde infty then
              uint 32 r === 0
            else
	      uint 32 r ==\midtilde mult }
\end{Verbatim}
Informally, the first post-condition states that the result is in canonical form, i.e. fully reduced.
The second post-condition states that, after the high bit of the $x$-coordinate of $P$ is masked (as per RFC7748\cite{rfc7748}), a ratio $(\ratio{X}{Z})$ representing the $x$-coordinate of ${[s]\cdot P}$ is computed using repeated application of Montgomery's formulas (where $[s]$ is the clamped value of $s$).
If $[s]\cdot P$ happens to be $\mathcal{O}$, the function writes a zero result; otherwise the result will be equivalent to $x_{[s]\cdot P}$.

Note that is not possible to distinguish the result $[s]\cdot P = \mathcal{O}$ and $x_{[s]P} = 0$.
However, for every point $P$ whose $y$-coordinate is not-zero, X25519 also does not
distinguish $P$ and $-P$; this specification elucidates that $\mathcal{O}$ and the point at the origin ($x=0, y=0$)
are similarly unified.

\section{Improved X25519 for AVR}
\label{sec:bugs}
Several small improvements were observed, which we confirmed by a formal proof. Two instructions
in the $128\times 128$-bit multiplication assembly routine could be removed with no impact on the formal proof, confirming they were
unnecessary. In \texttt{fe25519\_freeze}, the routine \texttt{bigint\_subp} is called twice to fully reduce an integer
mod $2^{255}-19$. We were able to verify that one call suffices, since in the current implementation it is always applied to a result that is already partially reduced.

\subsection{Memory Safety}
In \cite{vstte17}, several version of the Karatsuba implementations
could compute incorrect results if the memory locations used for storing input and output were
aliased, so we were naturally curious about aliasing in the X25519 implementation. 
We found that the prohibition on aliasing also applies to the $128$-bit and $256$-bit multiplication/squaring
routines,\footnote{As a peculiar exception: the $128$-bit squaring routine will function properly when reading from address $i$ and writing to address $i+8$}
and the \texttt{fe25519\_red} modular reduction function.
The modular addition/subtraction routines and \texttt{fe25519\_mul121666} were verified to be safe when used for in-place update operations.

The C code calls all these functions accordingly, so aliasing never becomes an issue.
We did add a \texttt{restrict} keyword to the function prototypes for which argument aliasing results in undefined behavior.

\subsection{Interrupt Safety}
\label{sec:interrupt}
The $256$-bit multiplication and squaring routines use function calls to the $128$-bit versions to compute their results, which
expect their arguments to be in memory. One of these calls multiplies an intermediate result
and so has to write this back to memory using the stack.

However, the original code did this by writing the data below the stack pointer.
This means that if the microcontroller is interrupted importunely (e.g. due to a timer or I/O event), and
an associated interrupt service routine needs this stack space for local variables, this data is clobbered.
The problem can be demonstrated by forcing an interrupt.

This problem was discovered during the modeling phase of verification, as our initial AVR model needed an extension to support direct access to the stack pointer, forcing us to consider the conditions under which this is allowed. We replaced the faulty code with code that moves the stack pointer using an idiomatic sequence\cite{avrlibc}, which we added to our model. Due to our formal proof, we were also able to see that in the $256\times 256$-bit multiplication some of the memory reserved for the final output was available for use as a temporary, reducing the amount of total stack space required by 32 bytes.

\section{Related Work}
Verified cryptography has gained much interest. In \cite{zinzin}, a verified library of elliptic curves written in F*
is presented. These provide the foundation for the C implementation of X25519 in the HACL* library\cite{hacl}: an implementation is created in a intermediate language Low*, verified against the F* specifications, and then
mechanically translated into C.
EverCrypt \cite{evercrypt} includes a similar C implementation, as well as an efficient implementation in x86-64 assembly code,
which is similarly generated, but using the Vale\cite{vale} tool. Vale is essentially a high-level assembly language with
support for
deductive reasoning, with a focus on cryptographic applications.
A similar X25519 implementation, now included in BoringSSL\cite{boringssl}, uses Coq\cite{coq} to generate efficient C code.
All these approaches involve \emph{generation} of \emph{new} implementations.

Efforts to verify \emph{existing} full implementations also exist.
In \cite{galois-saw}, an ECDSA implementation in Java is proven equivalent with a Cryptol\cite{cryptol} specification.
This is also a partially automated proof, requiring 1500 lines of annotation guiding the proof (in the form of SAWScript).
Compared to our approach, the Cryptol specification is less succinct---it actually is a complete, low-level implementation in its own right,
written in a functional language.

Recently, the X25519 implementation in TweetNaCl has been verified\cite{viguier} using Coq and VST\cite{vst}. This implementation was, however, designed with verification in mind. The proof states that TweetNaCl (when compiled with CompCert\cite{compcert}) correctly implements a scalar multiplication. Like \cite{zinzin}, the authors show this with respect to a formal \emph{mathematical} specification of elliptic curves.

Two efficient X25519 implementations written in 64-bit \texttt{qhasm} were partially verified by Chen et al\cite{cv25519}. Their approach is comparable to ours, in that they generate verification conditions which they solve using Boolector. However, where we use Why3 for this, they uses a custom approach, and report lengthier verification times. Their verification is partial, in the sense that they show that their Montgomery ladderstep implementation matches that of Algorithm~\ref{alg:ladderstep}, but don't verify the ladder itself. 
Similarly, Liu et al.\cite{liu-arithmetic} have verified several C routines of OpenSSL by compiling them to the LLVM intermediate representation, and translating that to the dedicated verification language \textsc{CryptoLine}.

\label{sec:related}

\section{Conclusion}
\label{sec:conclude}
To our knowledge, our result is the first to fully verify an existing high-speed implementation of X25519 scalar multiplication, and the first to present a verified implementation optimized for low-power devices.
We show correctness with respect to short formulas that are themselves proven correct in the literature\cite{monty87, bernstein-dh}.

Like \cite{viguier}, only general purpose, well-understood verification methods were used. 
Why3 in particular has an easy learning curve\cite{cserc19}. 
Our method for translating C and assembly code into WhyML is straight-forward, and the AVR model of Section~\ref{sec:avrmodel} can be validated, so trust in our results mainly resides with trusting the verification condition generation of Why3, the soundness of the automated provers, and the compilation-toolchain (C compiler, assembler and linker) used for producing AVR binaries.
The weakest link in this chain is definitely the use of automated provers: during our work we discovered a soundness error in Alt-Ergo 2.0, forcing us to preclude its use. 
We eagerly await the ability to perform proof reconstruction in Why3 using verified SMT solvers\cite{z3,bettersmt}.

We used a version of Why3 compatible with \cite{vstte17}. Newer versions are available, which in principle allow for an improved AVR model and specification. However, due to a change in the meaning of \emph{type invariants}, the versions
available to us generated inefficient SMT output for the verified multiplication routines of \cite{vstte17}. 
Since our use of type invariants can be avoided, we explored several alternatives,
but in the end chose to use the older version for time-efficiency reasons.

Our verification was performed in an amount of time that seems commensurate with the time it took the original implementers to
engineer the code.
Most time was spent on the multiplication routines in assembly code.
For the C code, the most time-consuming part was, in fact, finding the right abstraction level for a simple specification of the Montgomery ladder.

Due to our general purpose approach, our findings are encouraging for other low-level language applications.
In particular, due to the limitations of AVR, the code we encountered was quite long, and performed arithmetic
on many \emph{limbs} (32 instead of the more usual four or five). We expect our approach to work well for verifying the 32-bit ARM code in \cite{hutter}, requiring less time and with the possibility of some proof re-use.
We would also like to verify the compiler-generated assembly code of routines verified at a higher
level (such as in Section~\ref{sec:c}), by translating high-level specifications to the assembly level.
This would strengthen our result by removing the C compiler from the trusted code base.

\subsubsection*{Acknowledgments} 
The authors thank Beno\^it Viguier and Peter Schwabe for their advice, as well as the anonymous reviewers for their comments.
This material is based upon work supported by the Defense Advanced Research Projects Agency (DARPA) under Agreement No. HR.00112090028.
This work is part of the research programme `Sovereign' with project number 14319 which is (partly) financed by the Netherlands Organisation for Scientific Research (NWO).

\clearpage
\bibliographystyle{splncs04}
\bibliography{nfm21}

\iftrue
\appendix
\newpage

\section{Formal specification of X25519 Scalar Multiplication }
\begin{Verbatim}[frame=single,commandchars=\\\[\]]
type ratio = { x: int; z: int }
constant infty: ratio = {x=1;z=0}

constant p25519: int = pow2 255 - 19
predicate (===) (x y: int) = mod x p25519 = mod y p25519
predicate (\midtilde) (p q: ratio) = x p*z q === x q*z p
predicate (==\midtilde) (x:int) (xz: ratio) = xz \midtilde {x=x; z=1}

function add (m n mn: ratio): ratio
  = { x = 4*z mn*sqr(x m*x n - z m*z n); 
      z = 4*x mn*sqr(x m*z n - z m*x n) }

function double (n: ratio): ratio
  = { x = sqr(sqr(x n) - sqr(z n)); 
      z = 4*x n*z n * (sqr(x n) + 486662*x n*z n + sqr(z n)) }

function ladder (n: int) (p: ratio): (ratio, ratio)

axiom ladder_0: \emph[\footnotesize(*these axiomatic definitions are proven consistent*)]
  forall p.ladder 0 p = ({x=1; z=0}, p)

axiom ladder_even:
  forall p, n. n > 0 -> let (r0,r1) = ladder n p in 
    ladder (2*n) p = (double r0, add r1 r0 p)

axiom ladder_odd:
  forall p, n. n >= 0 -> let (r0,r1) = ladder n p in 
    ladder (2*n+1) p = (add r1 r0 p, double r1)

function scale (n: int) (m: int): ratio
  = let (r,_) = ladder n {x=m; z=1} in r

function clamp (x: int): int
  = mod x (pow2 254) + pow2 254 - mod x 8

val crypto_scalarmult_curve25519 (r s p: address_space)
  ensures { uint 32 r = mod (uint 32 r) p25519 }
  ensures { let xp   = mod (uint 32 p) (pow2 255) in
            let mult = scale (clamp (uint 32 s)) xp in
            if mult \midtilde infty then
              uint 32 r === 0
            else
              uint 32 r ==\midtilde mult }
\end{Verbatim}
\fi
\end{document}